\documentclass[onecolumn]{elsart3p}

\usepackage{graphics}
\usepackage{graphicx}
\usepackage{amssymb}

\usepackage[english,francais]{babel}

\newtheorem{e-proposition}[theorem]{Proposition}

\newtheorem{e-definition}[theorem]{Definition\rm}

\renewcommand{\theequation}{\arabic{equation}}
\newcommand{\du}{{\rm d}}
\newcommand{\be}{\begin{equation}}
\newcommand{\ee}{\end{equation}}
\newcommand{\ben}{\begin{eqnarray}}
\newcommand{\een}{\end{eqnarray}}
\newcommand{\beq}{\begin{equation}}
\newcommand{\eeq}{\end{equation}}

\newcommand{\B}{\mathrm{B}}
\newcommand{\NB}{\mathrm{NB}}

\def\og{\leavevmode\raise.3ex\hbox{$\scriptscriptstyle\langle\!\langle$~}}
\def\fg{\leavevmode\raise.3ex\hbox{~$\!\scriptscriptstyle\,\rangle\!\rangle$}}

\DeclareGraphicsExtensions{.png,.pdf,.jpg,.jpeg}      % for pdflatex

\begin{document}
%  You can place here the title of the dossier, if you know it,
%     firstly in English, then in French
\centerline{Quasicrystals / Les quasicristaux}
\begin{frontmatter}

% Title, authors and addresses

% use the thanksref command within \title, \author or \address for footnotes;
% use the ead command for the email address,
% and the form \ead[url] for the home page:
% \title{Title\thanksref{label1}}
% \thanks[label1]{}
% \author{Name\thanksref{label2}}
% \ead{email address}
% \ead[url]{home page}
% \thanks[label2]{}
% \address{Address\thanksref{label3}}
% \thanks[label3]{}
\selectlanguage{english}
%\title{Anomalous electronic transport in quasicrystals and related complex metallic alloys}

\title{Anomalous electronic transport in quasicrystals and related complex metallic alloys}

% use optional labels to link authors explicitly to addresses:
% \author[label1,label2]{}
% \address[label1]{}
% \address[label2]{}
% If all authors are at the same address, the [label1] can be suppressed

\selectlanguage{english}
\author[authorlabel1]{Guy Trambly de Laissardi\`ere},
\ead{guy.trambly@u-cergy.fr}
\author[authorlabel2]{Didier Mayou}
\ead{didier.mayou@grenoble.cnrs.fr}

\address[authorlabel1]{Laboratoire de Physique Th\'eorique et Mod\'elisation, Universit\'e de Cergy-Pontoise - CNRS UMR 8089, F-95302 Cergy-Pontoise, France.}
\address[authorlabel2]{Universit\'e Grenoble Alpes, Institut NEEL, F-38042 Grenoble, France \\
CNRS, Institut NEEL, F-38042 Grenoble, France}

\begin{abstract}
We analyze the transport properties in approximants of quasicrystals $\alpha$-AlMnSi, $1/1$-AlCuFe and for the complex metallic phase $\lambda$-AlMn. These phases presents strong analogies in their local atomic structures and are related to existing quasicrystalline phases. Experimentally they present unusual transport properties with low conductivities and a mix of metallic-like and insulating-like characteristics. We compute the band structure and the quantum diffusion in the perfect structure without disorder and introduce simple approximations that allow to treat the effect of disorder. Our results demonstrate that the standard Bloch-Boltzmann theory is not applicable to these intermetallic phases. Indeed  their dispersion relation are flat indicating small band velocities and corrections to quantum diffusion  that are not taken into account in the semi-classical Bloch-Boltzmann scheme become dominant. We call this regime the small velocity regime. A simple Relaxation Time Approximation to treat the effect of disorder allows us to reproduce the main experimental facts on conductivity qualitatively and even quantitatively.

{\it To cite this article: Guy Trambly de Laissardi\`ere, D. Mayou, 
C. R. Physique XXXX (20XX).}

\vskip 0.5\baselineskip

\selectlanguage{francais}
\noindent{\bf R\'esum\'e}
\vskip 0.5\baselineskip
\noindent
{\bf Transport \'electronique anormal dans les quasicristaux et les alliages m\'etalliques complexes reli\'es}.\\
Nous analysons les propri\'et\'es de transport \'electronique dans les approximants de quasicristaux $\alpha$-AlMnSi, $1/1$-AlCuFe et pour la phase complexe  reli\'ee $\lambda$-AlMn. Ces phases pr\'esentent de fortes analogies de leurs structures atomiques locales et sont reli\'ees \`a des phases quasicrystallines existantes. Exp\'erimentallement elles pr\'esentent des propri\'et\'es de transport inhabituelles avec une faible conductivit\'e et un m\'elange de propri\'et\'es de type m\'etallique et de type isolant. Nous calculons la structure de bande et la diffusion quantique de la structure parfaite et introduisons une approximation simple qui permet de traiter l'effet du d\'esordre. Nos r\'esultats d\'emontrent que la th\'eorie standard de Bloch-Boltzmann n'est pas applicable \`a ces syst\`emes interm\'etalliques. En effet leurs relations de dispersion sont plates indiquant une faible vitesse de bande et les corrections \`a la diffusion quantique qui ne sont pas prises en compte par la th\'eorie semi-classique deviennent dominantes. Nous appelons ce r\'egime le r\'egime de faible vitesse.  Une simple approximation de temps de relaxation pour traiter l'effet du d\'esordre permet de reproduire les principaux r\'esultats exp\'erimentaux sur la conductivit\'e qualitativement et m\^eme quantitativement.

{\it Pour citer cet article~: Guy Trambly de Laissardi\`ere, D. Mayou, 
C. R. Physique XXXX (20XX).}

%Now keywords/mots-cles
 \vskip 0.5\baselineskip
\noindent{\small{\it Mots-cl\'es~:} calcul ab-initio~; quasicristaux~; alliages interm\'etalliques complexes~;
transport \'electronique}}
\end{abstract}
\end{frontmatter}

\selectlanguage{english}
% main text
\section{Introduction}
\label{SecIntro}

Immediately after the discovery by Shechtman et al. \cite{shechtman84} of quasiperiodic intermetallics one major question was raised about the physical properties of  phases with this new type of order. In particular, one expected that the electronic and thermal properties could be deeply affected \cite{Poon92,Berger93c,Berger94,Grenet00_Aussois}. Indeed the description of electrons or phonons in periodic phases rests on the Bloch theorem which cannot be applied to a quasiperiodic structure. Within a decade a series of new quasiperiodic phases and approximant were discovered and intensively studied. These investigations learned us that indeed the electrons and the phonons properties could be deeply affected by this new type of order.

The first quasiperiodic alloys where metastable and contained many structural defects. As a consequence they had conduction properties similar to those of amorphous metals with resistivities in the range 100--500\,$\mu\Omega$cm. 
%In 1986 the first stable icosahedral phase was discovered in AlLiCu. This phase was still defective and although  its resistivity was higher (800\,$\mu\Omega$cm) it was still comparable to that of amorphous metals. 
The real breakthrough came with the discovery of the stable AlCuFe icosahedral phase, having a high structural order. The resistivity of these very well ordered systems where very high, of the order of 10\,000 $\mu\Omega$cm \cite{Klein91,Mayou93},  which gave a considerable interest in their conduction properties.  Within a few years several important electronic characteristics of these phases  were experimentally demonstrated.  The density of states in AlCuFe was smaller than in Al, about one third of that of pure Al, but still largely metallic. The conductivity presented a set of characteristics that were either  that of semi-conductors or that of normal metals. In particular  weak-localization effects were observed that are typical of amorphous metals. Yet the conductivity was increasing with the number of defects just as in semi-conductors.  
Optical measurements showed that the Drude peak, characteristic of normal metals,  was absent. 
In 1993 another breakthrough was the discovery of  AlPdRe  which had resistivities in the range of  $10^6\,\mu\Omega$cm \cite{Pierce93_science,Berger93,Akiyama93}. 
This system gave the possibility of studying a metal-insulator transition in a quasiperiodic phase.  There are still many questions concerning electronic transport in AlPdRe phases. 
%One difficulty concerns the homogeneity and the quality of samples which are  crucial for transport properties but are difficult to determine exactly.  

Since the discovery the view of the role of quasiperiodic order has evolved. 
On  one hand, the long-range quasiperiodic order
can induce electronic states  neither localized nor
extended, called 
\textit{``critical states''} (see Ref. \cite{Sire94} and Refs. therein).
On the other hand,
for  electronic or phonon properties of most known alloys  it appears that the medium range order, on one or a few nanometers, is the real length scale that determines properties. 
This observation has lead the scientific  community to adopt a larger point of view and consider quasicrystals as an example of a larger class. This  class of Complex Metallic Alloys contains  quasicrystals, approximants and alloys with large and complex unit cells with possibly hundreds of atoms in the unit cell. 

In this article we study ``how electrons propagate'' 
in aluminum based quasicrystals, approximants or complex metallic alloys with  structure related to quasiperiodicity. 
The main objective is to show that the non standard  conduction properties of some quasicrystals and related complex metallic alloys result from purely quantum effects and cannot be interpreted through the semi-classical theory of transport. This is of great importance since the semi-classical Bloch-Boltzmann theory is at the heart of our understanding of conduction in solids, 
ranging from metals to semi-conductors.  

This new type of quantum transport is related to the specific  propagation mode of electrons in these  systems. Indeed in quasicrystals and related complex phases the quantum diffusion law deviates from the standard ballistic law characteristic of  perfect crystals in two possible ways. 
In a perfect quasicrystal the large time diffusion law is a power law  instead of a ballistic one in perfect crystals. In a complex crystal the diffusion law is always ballistic at large time but  it can deviate strongly from the ballistic law at sufficiently small times. It is this specific character that provides a basis for the interpretation of the strange conduction properties of AlCuFe, AlPdMn and probably also for those of AlPdRe.

This paper is organized as follows.  In section \ref{SecMethod},
the formalism of linear response for conductivity (Kubo formalism) is presented in terms of the quantum diffusion. The  numerical method to calculate numerically quantum transport in actual phases is described briefly. Then we present results in section  \ref{abinitioSec} for two approximants phases $\alpha$-AlMnSi, $1/1$-AlCuFe and for the complex metallic phase $\lambda$-AlMn. These results show that  the Boltzmann approach is no more valid in these systems. This is because the electron velocity is small (flat bands) and wave packets have large spatial extension. We call this regime of transport the small velocity regime. 
In section \ref{SecMIT} we propose a simple phase diagram, at zero temperature, for the Anderson Metal-Insulator transition in  phases within the small velocity regime. As we show the small velocity regime deeply influences the occurrence of the Anderson transition in the presence of static disorder. In standard systems the Anderson transition always occurs when the disorder increases whereas here the behavior is more complex. In the conclusion (section \ref{Conclusion}) we briefly summarize our main findings and discuss some open questions as well as the connection to other systems.

\section{Formalism for quantum diffusion calculation}
\label{SecMethod}

\subsection{Quantum diffusion and conductivity}

The present study relies upon the evaluation of the Kubo-Greenwood conductivity using the Einstein relation between the conductivity and the quantum diffusion
\cite{Mayou88,Mayou95,Roche97,Roche99,Triozon02,Mayou00,Mayou07_revue,Darancet10}.
Central quantities are the velocity correlation function of
states of energy $E$ at time $t$,
\begin{eqnarray}
C(E,t) = \Big\langle \hat{V}_x(t)\hat{V}_x(0) + \hat{V}_x(0)\hat{V}_x(t) \Big\rangle_E
= 2\,{\rm Re}\, \Big\langle \hat{V}_x(t)\hat{V}_x(0) \Big\rangle_E \, ,
\label{EqAutocorVit}
\end{eqnarray}
and the average
square spreading (quantum diffusion) of states of energy $E$ at time $t$
along the $x$ direction,
\begin{equation}
X^{2}(E,t)= \left \langle \left( \hat{X}(t)- \hat{X}(0)^{2} \right) \right \rangle_{E}.
\label{XET}
\end{equation}
%where $X(t)$ is the position operator along the $x$-direction in the Heisenberg representation and $\langle %~\rangle_{E}$ means an average over states with energy $E$, usually the Fermi energy. 
In equations (\ref{EqAutocorVit}) and  (\ref{XET}), 
$\left \langle ... \right \rangle_{E}$ is the average on states with energy $E$,  
Re$\,{{A}}$ is the real part of ${{A}}$,
$\hat{V}_x(t)$ and $\hat{X}(t)$ are the Heisenberg representation of the velocity operator $\hat{V}_x$ 
and the position operator $\hat{X}$
along $x$ direction at time $t$,
\begin{eqnarray}
\hat{V}_x = \frac{1}{i \hbar}~ \Big[ \hat{X} , \hat{H} \Big].
\end{eqnarray}
$C(E,t)$ is related to quantum
diffusion by the relation \cite{Mayou00},
\begin{eqnarray}
\frac{\rm d}{{\rm d} t} \Big(X^2(E,t) \Big) = \int_0^{t}C(E,t'){\rm d} t'.
\label{EqX2}
\end{eqnarray}
From Kubo-Greenwood formula, 
the conductivity is given by the Einstein relation,
\begin{eqnarray}
\sigma(E_{\rm F}) = e^2 n(E_{\rm F}) D(E_{\rm F}),
\label{EinsteinRelation}
\end{eqnarray}
where $e$ is the electron charge, $E_{\rm F}$ the Fermi energy,  $n$ the density of states and $D$ the diffusivity related 
to the square spreading by the relation \cite{Mayou00,Mayou07_revue},
\begin{eqnarray}
D(E_{\rm F}) = \frac{1}{2} \, \lim_{t\rightarrow \infty} \frac {{\rm d}}{{\rm d} t} \,  X^2(E_{\rm F},t).
\label{EqDiffusivity}
\end{eqnarray}

\subsection{Conductivity in perfect periodic systems}
\label{subsecPerfect}

In crystals, these quantities can be decomposed 
in a ballistic contribution (Boltzmann term) and a 
non-ballistic contributions (non-Bolzmann term):
\begin{equation}
C(E,t)= 2~ V_{\mathrm{B}}(E)^2  + C_{\NB}(E,t) 
\label{Eq_autocorVitesse}
\end{equation}
and after equation  (\ref{EqX2})
\begin{equation}
X^2(E,t) = V_{\mathrm{B}}(E)^2 t^2 + X_{\mathrm{NB}}^2(E,t),
\label{Eq_DeltaX2}
\end{equation}
where $V_\mathrm{B}(E)$ is 
the Boltzmann velocity at energy $E$,
\be
V_{\mathrm{B}}(E)^2 =  \left\langle 
|\langle n\vec k | \hat{V}_x | n\vec k \rangle |^2
\right\rangle_{E_n=E}.
\label{equationVB}
\ee
$V_{\mathrm{B}}(E)$ is also the average band velocity at the energy $E$ in $x$ direction, since the band velocity is given by 
 \be
\frac{1}{\hbar} \frac{\partial E_n(\vec k)}{\partial k_x}=\langle n\vec k | \hat{V}_x | n\vec k \rangle
\ee
where $E_n$ is the energy of the eigenstate  $| n\vec k \rangle$ at wave vector $\vec k$.
In (\ref{Eq_autocorVitesse}) and (\ref{Eq_DeltaX2}), the ballistic terms 
$C_{\mathrm{B}} =2 V_{\mathrm{B}}(E)^2$ and
$ X_{\mathrm{B}} = V_{\mathrm{B}}(E)^2 t^2$ 
are due to intra-band contributions.
And the non-ballistic terms $C_{\NB}(E,t)$, $ X_{\mathrm{NB}}^2(E,t)$ 
are due to the inter-band contributions:
\begin{equation}
X^2_{\mathrm{NB}}(E_{\mathrm{F}},t) = 2 \hbar^2
~\Bigg\langle
\sum_{m \,(m\neq n)}
\frac{1 - \cos\Big((E_n-E_m)\frac{t}{\hbar} \Big)}{(E_n-E_m)^2} 
\Big| \langle n\vec k | \hat{V}_x | m\vec k \rangle \Big|^2
\Bigg\rangle_{E_n=E_{\mathrm{F}}}.
\label{Calcul_DeltaX2}
\end{equation}
$ X_{\rm NB}^2(E,t)$ is the average spreading of the state
for large time $t$ it oscillates (see next section).
From its maximum value the length $L_{wp}(E)$ is defined:
\beq
 X_{\rm NB}^2(E,t) \leq L_{wp}(E)
\eeq
$L_{wp}(E)$ represents 
the average expansion of wave packet eigenstates
at energy $E$. 
Therefore a small $L_{wp}(E)$ value is espected for confined states by atomic clusters 
\cite{Janot94,GuyPRB97,GuyICQ6}. 
In the Boltzmann theory for electronic transport, the non-Boltzmann contributions are neglected. 
This is a resonnable 
assumption for rather simple metallic phases, but non-Boltzmann terms are essentiel to understand complex 
metallic alloys such as quasicrystals and related phases. 

In practice, from self-consistent LMTO eigenstates or Tight-Binding eigenstates,
we compute the velocity correlation function  
$C(E,t)$ and $X(E,t)$ for crystals.
In equations ($\ref{EqAutocorVit}$), (\ref{XET}), (\ref{equationVB}) and ($\ref{Calcul_DeltaX2}$) the average
is obtained by taking the eigenstates for each
$\vec{k}$ vector with and energy $E_n(\vec{k})$ such as
%$E-\Delta E /2<E_n(\vec{k})<E+\Delta E/2$.
\begin{eqnarray}
E-\frac{1}{2}\delta E<E_n(\vec{k})<E+\frac{1}{2}\delta E.
\end{eqnarray}
$\delta E$ is the energy resolution of the calculation. 
When number $N_k$ of $\vec{k}$ vectors in the first Brillouin zone is too small, the calculated
quantities are sensitive to $N_k$.
Therefore $N_k$ is increased
until results do not depend significantly on $N_k$.
We use $\delta E = 0.01$\,eV;
$N_k = 32^3$ for 1/1-AlCuFe and $\alpha$-AlMnSi, 
and $N_k = 8\times8\times16$ for $\lambda$-AlMn.

\subsection{Conductivity in system with defects}
\label{SecRes}

The effect of the elastic scattering (static defects) and/or the inelastic scattering (electron-phonon, electron-electron) can be treated in a phenomenological way in the scheme of the  Relaxation Time Approximation (RTA) \cite{Mayou00,Mayou07_revue}.
We introduce a scattering time $\tau$, beyond which the propagation becomes diffusive due to the destruction  of coherence through  scattering by defects. 
%This relaxation time approximation (RTA) has been used succesfully to compute \cite{Trambly06} 
%conductivity in approximants of quasicrystal where quantum diffusion and localization effect play a %essential role \cite{Berger93,Belin93,Trambly05}. 
Following previous works \cite{Mayou00,Mayou07_revue,PRL06,Ciuchi11}, we assume that the velocity correlation function $C_{s}(E,t)$ of the system with  scatterers (defects) is given by,
\begin{eqnarray}
C_{s}(E,t) ~\simeq ~ C(E,t)\, {\rm e}^{-|t|/\tau} \, ,
\label{RTA_C}
\end{eqnarray}
where $C(E,t)$ is the velocity correlation of the system 
without defects. 
%Here the  scattering time $\tau$ is the cutoff time of the weak 
%localization effects also called dephasing time.
The propagation given by this formalism is unaffected by scattering at short times ($t < \tau$) and diffusive at long times ($t> \tau$) as it must be.  
Using the $t=0$ conditions, $X^2(E,t=0)=0$ and $\frac{{\rm d}}{{\rm d}t}X^2(E,t=0)=0$, 
and performing two integrations by part,
we obtain from equations
 (\ref{EqX2}), (\ref{EinsteinRelation}), (\ref{EqDiffusivity}) and (\ref{RTA_C}), \cite{Mayou07_revue}
\begin{eqnarray}
\sigma(E_{\rm F},\tau)&=& e^{2} n(E_{\rm F})D(E_{\rm F},\tau) \label{EquationEinstein} \, ,\\  
D(E_{\rm F},\tau)&=&\frac{L^{2}(E_{\rm F},\tau)}{2 \tau} \, ,\\
L^{2}(E_{\rm F},\tau)&=&\frac{1}{\tau} \int_0^\infty \! X^{2}(E_{\rm F},t)\,{\rm e}^{-t/\tau} \, \mathrm{d}t \, ,
\label{EquationLi}
\end{eqnarray}
where 
%$E_{\rm F}$ is the Fermi energy, $n(E_{\rm F})$ the density of states, $\tau_{i}$ the inelastic scattering time %and 
$L(E_{\rm F},\tau)$  the  mean-free path
and $D(E_{\rm F},\tau)$ the diffusivity. 
$X^{2}(E,t)$ is calculated for the system without defect (section \ref{subsecPerfect}). 
The above equations treat the scattering in a way that is equivalent to the standard approximation in mesoscopic physics. 
Indeed, in the presence of scattering,  it is usually assumed that,
$L(E_{\rm F}) \simeq \sqrt{X^{2}(E_{\rm F},\tau)}$, thus 
the conductivity is given by the Einstein formula with a diffusivity   
$D(E_{\rm F},\tau) \simeq X^{2}(E_{\rm F},\tau)/(2 \tau)$ \cite{Lee85},  
which is essentially equivalent to the above equations.

In periodic systems, 
the dc-diffusivity at energy $E$ is given by 
 \ben
D(E) = \frac{1}{2} \, 
\int_0^{+\infty} \e^{-t/\tau} C(E,t) {\du}t = D_{\B}(E)+D_{\NB}(E),
\label{D(C0)p}
\een
where Boltzmann diffusivity is $D_{\B}(E) = V_B^2(E) \tau$, 
and the non-Boltzmann term is
\ben
D_{\NB}(E) = \frac{1}{2} \,  
 \frac{1}{\tau^2}
\int_0^{+\infty} \e^{-t/\tau}  X_{\NB}^2(E,t) {\du}t.
\label{DNB21}
\een
%As shown figure \ref{Fig_DOS_VB_Dif_Sig}.c,
%$D_{\mathrm{NB}}$ is almost independent
%on $E$, whereas the
%$D_{\mathrm{B}}$ values depend strongly on $E$,
%as $V_{\B}$ value depends on $E$.

\section{Ab initio transport properties in approximants and complex metallic alloys}
\label{abinitioSec}

\subsection{Atomic structures}

To present the quantum diffusion in approximants of quasicrystals 
and complex metallic alloys related to quasiperiodicity we consider three phases:
the $\alpha$-AlMnSi approximant, a model for AlCuFeSi 1/1 cubic approximant
and the complex metallic phase   $\lambda$-AlMn.
Results for $\lambda$-AlMn are new and are compared to previous results for $\alpha$-AlMnSi and  1/1 AlCuFeSi that have already been presented in Ref. \cite{PRL06,Trambly08}.

For the  $\alpha$-AlMnSi phase, 
we use the experimental atomic structure
\cite{Sugiyama98}
with the Si positions proposed by Ref.
\cite{Zijlstra03} for the composition
$\alpha$-Al$_{69.6}$Si$_{13.0}$Mn$_{17.4}$.
This phase contains 138 atoms in a cubic unit cell:
96 Al atoms, 18 Si atoms, and 24 Mn atoms.

V. Simonet {\it et al.}~\cite{Simonet05_AlCuFe}
refined experimentally the atomic structure and the chemical
decoration of Al--Cu--Fe--Si 1/1 cubic approximants.
The authors give a revised description of the structure of  
$\alpha'$-$\rm Al_{71.7}Si_7Cu_{3.8}Fe_{17.5}$ phases
and  $\alpha$-$\rm Al_{55}Si_7Cu_{22.5}Fe_{12.5}$ phase.
$\alpha'$-phase has a chemical
decoration similar to that 
of $\alpha$-Al--Mn--Si, 
whereas the structure and the composition of the $\alpha$-phase is 
different.
It is characterized by several Wyckoff sites with mixed occupancy between
Al/Cu, Al/Fe and Cu/Fe.
As an example, we used
this  structure to calculate the ab-initio electronic structure for
phase with the composition $1/1$-Al$_{56.1}$Cu$_{34.5}$Fe$_{9.4}$ i.e.
$\rm Al_{78}Cu_{48}Fe_{13}$ in a cubic unit cell.

The complex $\rm \lambda$-$\rm Al_{4.6}Mn$ phase~\cite{Kreiner97} crystallizes in a
large hexagonal structure $\rm P6_3/mmc$ with a unit cell containing about
590 atoms. The structure of $\rm \lambda$-$\rm Al_{4.6}Mn$ phase is closely related to the 
hexagonal $\rm \mu$-$\rm Al_{4.12}Mn$ \cite{Shoemaker89}.
These phases are not approximant of quasicrystal but their local environment are related strongly 
to local and medium range order induce by the quasiperidicity \cite{Kreiner97,Kreiner95}.
We calculated electronic structure of  $\rm \lambda$ phase from LMTO
by using an atomic structure with minor modifications \cite{Duc03} from the
experimental structure to avoid mixed occupied
sites. The same modification has been done to study  $\rm \mu$-$\rm Al_{4.12}Mn$ \cite{Duc03}.
The structure used to calculate the electronic properties has the composition
$\rm Al_{483}Mn_{104}$ in a cubic unit cell.

\subsection{Density of states}

\begin{figure*}%[]
\begin{center}

\includegraphics[width=0.4\textwidth]{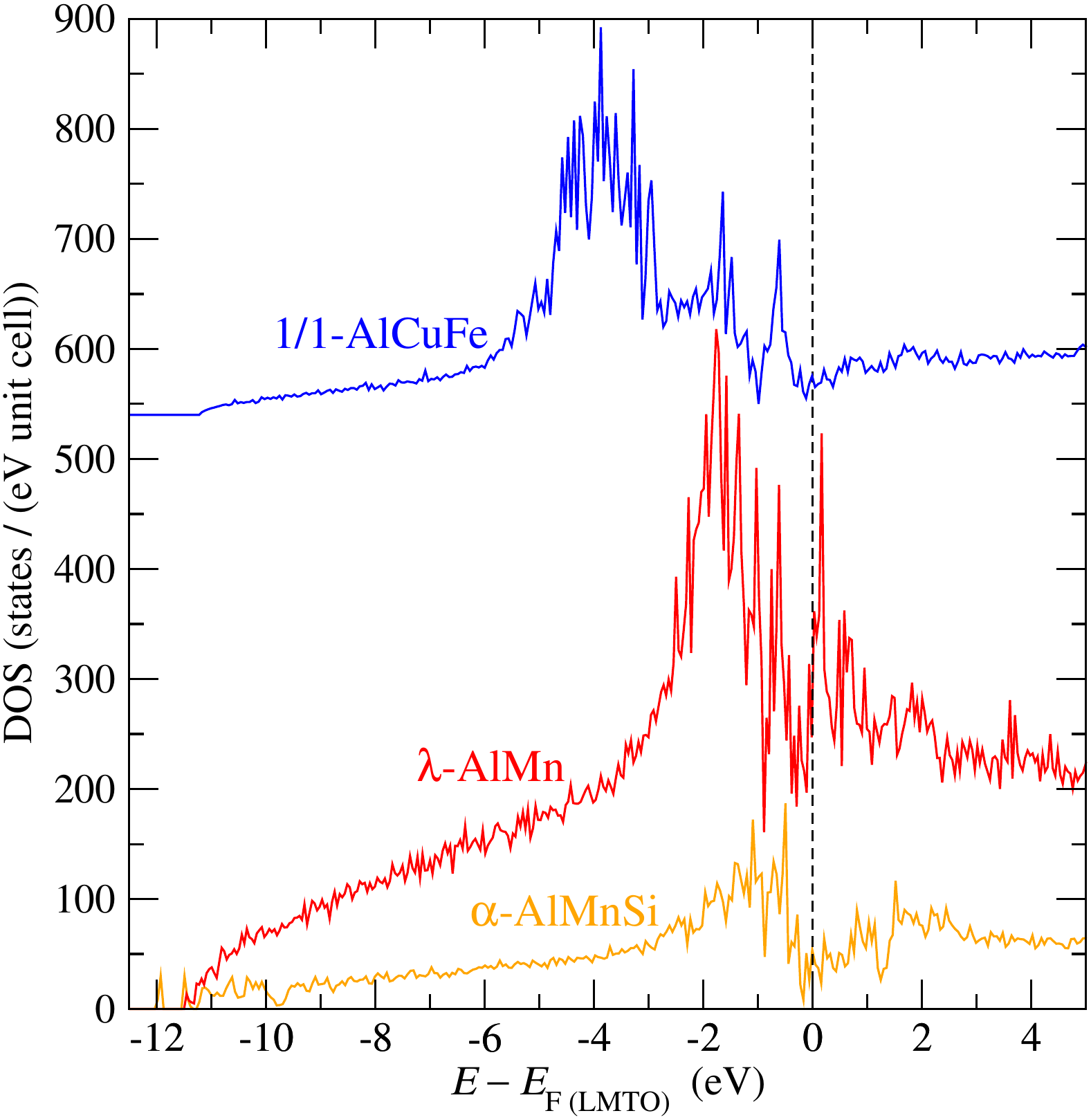}

\caption{\label{Fig_AlCuFe_alpha_lambda_DOS}
LMTO Density of states (DOS), $n(E)$,
in 
$1/1$-Al$_{56.1}$Cu$_{34.5}$Fe$_{9.4}$, 
$\alpha$-Al$_{69.6}$Mn$_{17.4}$Si$_{13.0}$ and 
$\lambda$-Al$_{4.6}$Mn.
}
\end{center}
\end{figure*}

\begin{figure*}%[]
\begin{center}

\includegraphics[width=0.45\textwidth]{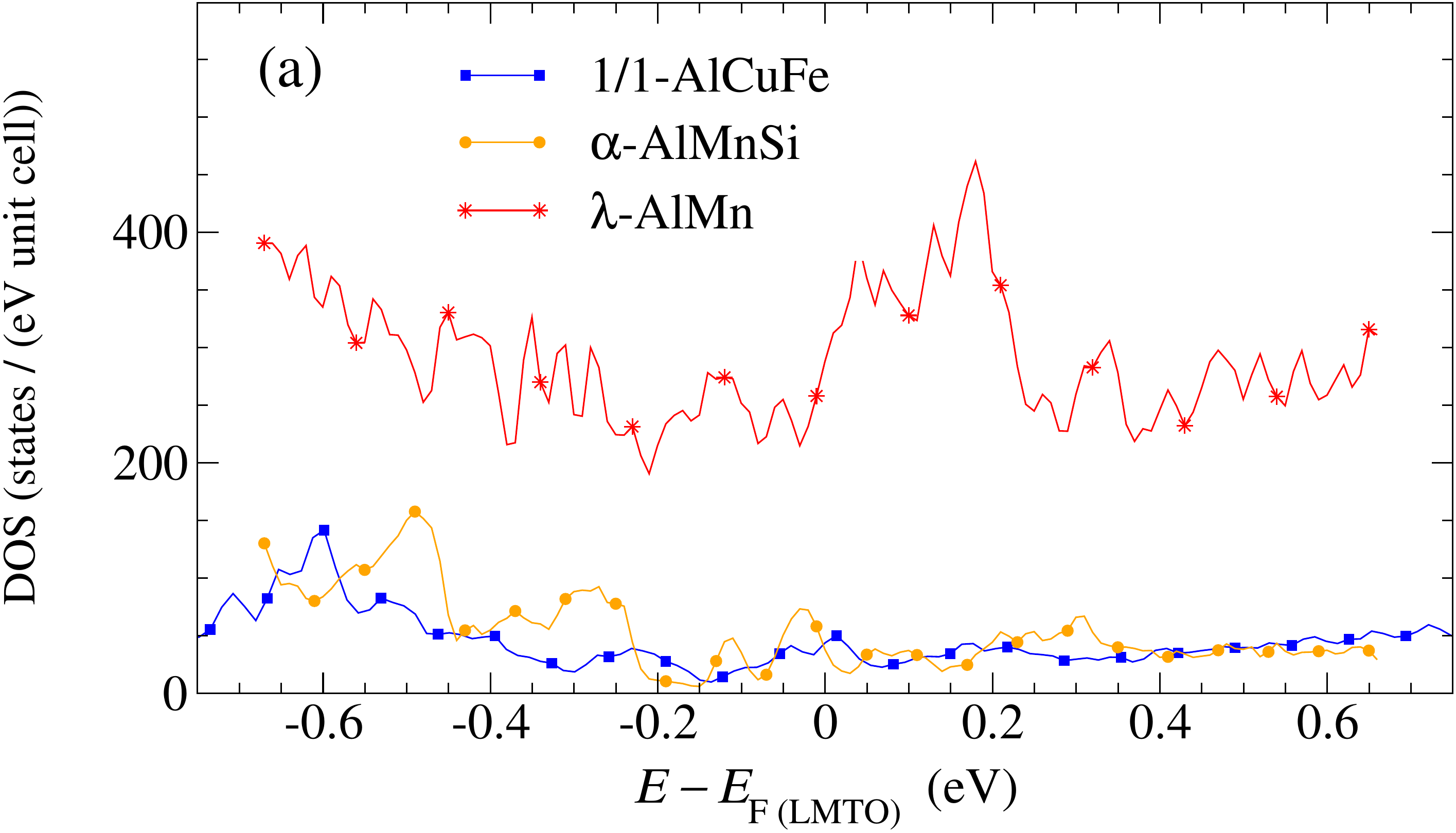}

\vskip .5cm
\includegraphics[width=0.45\textwidth]{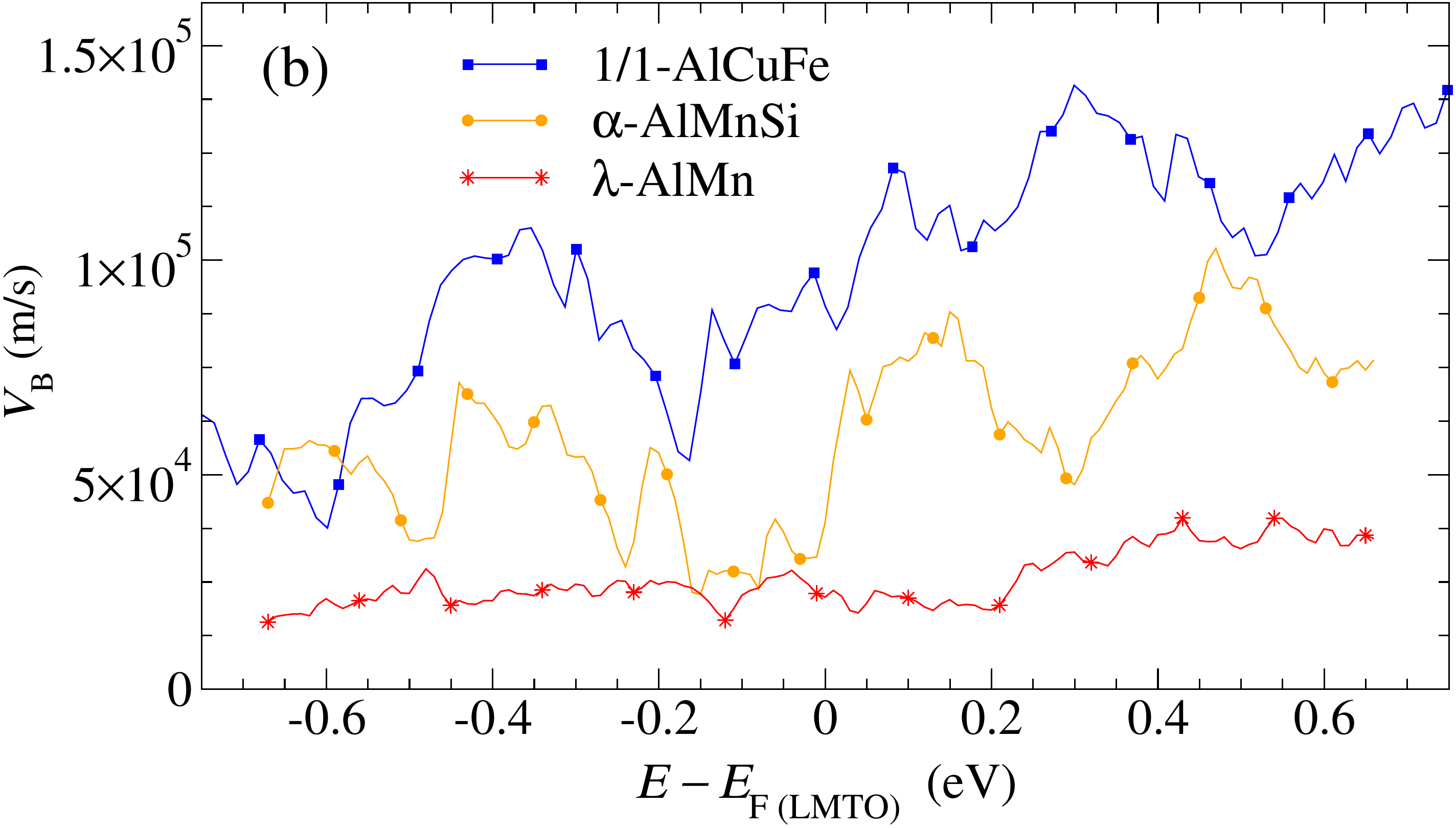}

\caption{\label{Fig_AlCuFe_alpha_lambda_DOS_VB}
LMTO (a) DOS and (b) Boltzmann  velocity 
in 
$1/1$-Al$_{56.1}$Cu$_{34.5}$Fe$_{9.4}$, 
$\alpha$-Al$_{69.6}$Mn$_{17.4}$Si$_{13.0}$ and 
$\lambda$-Al$_{4.6}$Mn, arround $E_{\rm F (LMTO)}$.
}
\end{center}
\end{figure*}

In the framework of the density functional approximation, the ab initio electronic structure of the studied phases is computed by using the LMTO method \cite{Andersen75,andersen95}.
In figures \ref{Fig_AlCuFe_alpha_lambda_DOS} and \ref{Fig_AlCuFe_alpha_lambda_DOS_VB}a, 
the non magnetic total density of states 
DOS, $n(E)$, of studied phases 
are presented. A pseudogap near $E_{\rm F}$ is clearly seen, its weight is about 200\,meV or more.
Following the Hume-Rothery condition,
the stabilization is obtained when the Fermi sphere matches  a pseudo-Brillouin zone
(also called Jones zone) (see Refs. \cite{Mizutani10,PMS05} and Refs. therein).
This condition is reached for  Fermi level
$E_{\mathrm{F}}$  in proximity of the minimum of the pseudogap.
It is well know that this pseudogap is due to the diffraction by Bragg planes 
of the pseudo-Brillouin zone,
but this mechanism is strongly increased by the $sp$-$d$ hybridisation 
between Al,Si $sp$ states and transition metal $d$ orbitals  \cite{Friedel87,NguyenManh87,dankhazi93,Belin93,Mayou93b,GuyEuro93,GuyPRB95,GuyPRBexp95,Krajvci95,simonet98,Hippert99,GuyPRL00,Guy03,PMS05}.
It must be noted that the diffraction by Bragg planes, leading to pseudogap in the DOS, can be understood in terms of 
oscillating pair potential interactions in the real space due to Friedel oscillations of the charge density \cite{Guy04_ICQ8,icq10}.

As shown first by T. Fujiwara \cite{Fujiwara89,Fujiwara93}, the DOS is also characterized by the presence of fine peaks, called
{\it ``spiky peaks''}. Their width is about $10-100$\,meV.
In approximants, they are a consequence of flat bands, $E_n(\vec k)$, in the reciprocal space and they show a new kind of confinement 
of electrons by the local and medium range atomic order. 
Indeed numerical calculations have shown that atomic clusters, with typical size equal to 20\,$\rm \AA$ or more, 
can confine electrons by forming {\it ``Cluster Virtual Bound States''} \cite{GuyPRB97,GuyICQ6,icq10}.
The existence or not of spiky peaks in the DOS of actual approximants and quasicrystals has been much debated experimentally and 
theoretically (see Refs. cited in \cite{icq10}). More recently low-temperature scanning tunneling 
spectroscopy \cite{Widmer06,Widmer09,Mader10,Mader13} confirmed the presence of fine peaks in the DOS of surfaces of 
 $\mu$-AlMn phases, icosahedral AlPdMn and decagonal Al–Ni–Co.

The Boltzmann velocity (intra-band velocity) $V_{\B}$ calculated from equation (\ref{equationVB})
is shown on figure  \ref{Fig_AlCuFe_alpha_lambda_DOS_VB}b.
These results are similar to the original work of  T. Fujiwara et al.
\cite{Fujiwara93,GuyPRB94_AlCuFe,GuyPRBAlCuCo} for approximants.
$V_{\B}$ in small approximants 1$/$1-AlCuFe and $\alpha$-AlMnSi varies very rapidly with a small variation of $E$, which shows 
the crucial effect of the chemical composition on transport properties.
The minimum value of $V_{\B}(E)$ is about $\rm 2 \times 10^6\,cm.s^{-1}$, 
whereas in simple
crystals Al (f.c.c.) and  cubic $\rm Al_{12}Mn$:
$V_{\rm B}= 9 \times 10^{7}$ and $\rm 4 \times 10^{7}\,cm.s^{-1}$, 
respectively \cite{ICQ9}.
For $\lambda$-AlMn, the reduction for $V_{\rm B}$ with respect to usual intermetallic alloys is even
stronger.

\subsection{Mean square spreading}

\begin{figure*}%[]
\begin{center}

\includegraphics[width=0.6\textwidth]{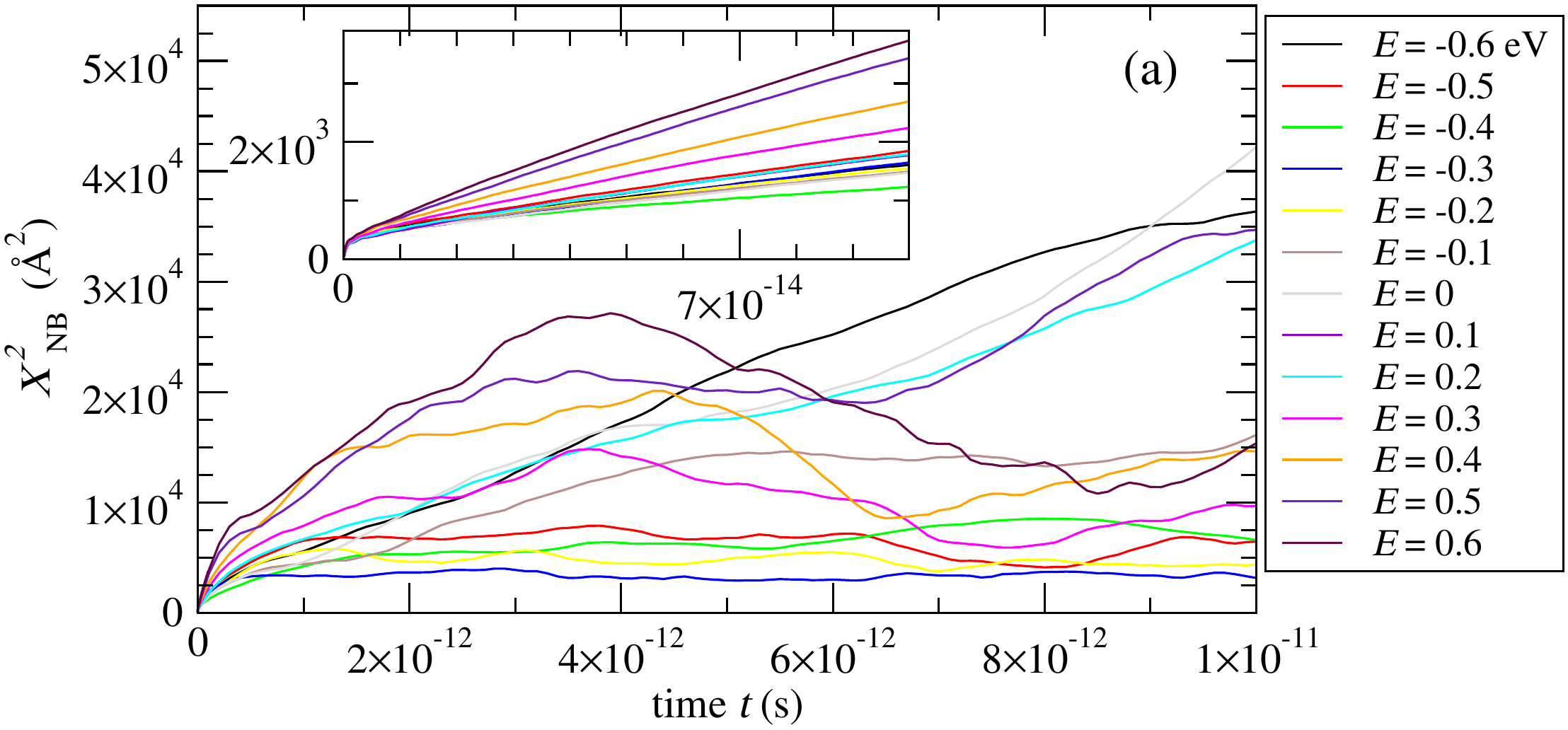}

\vskip .5cm
\includegraphics[width=0.62\textwidth]{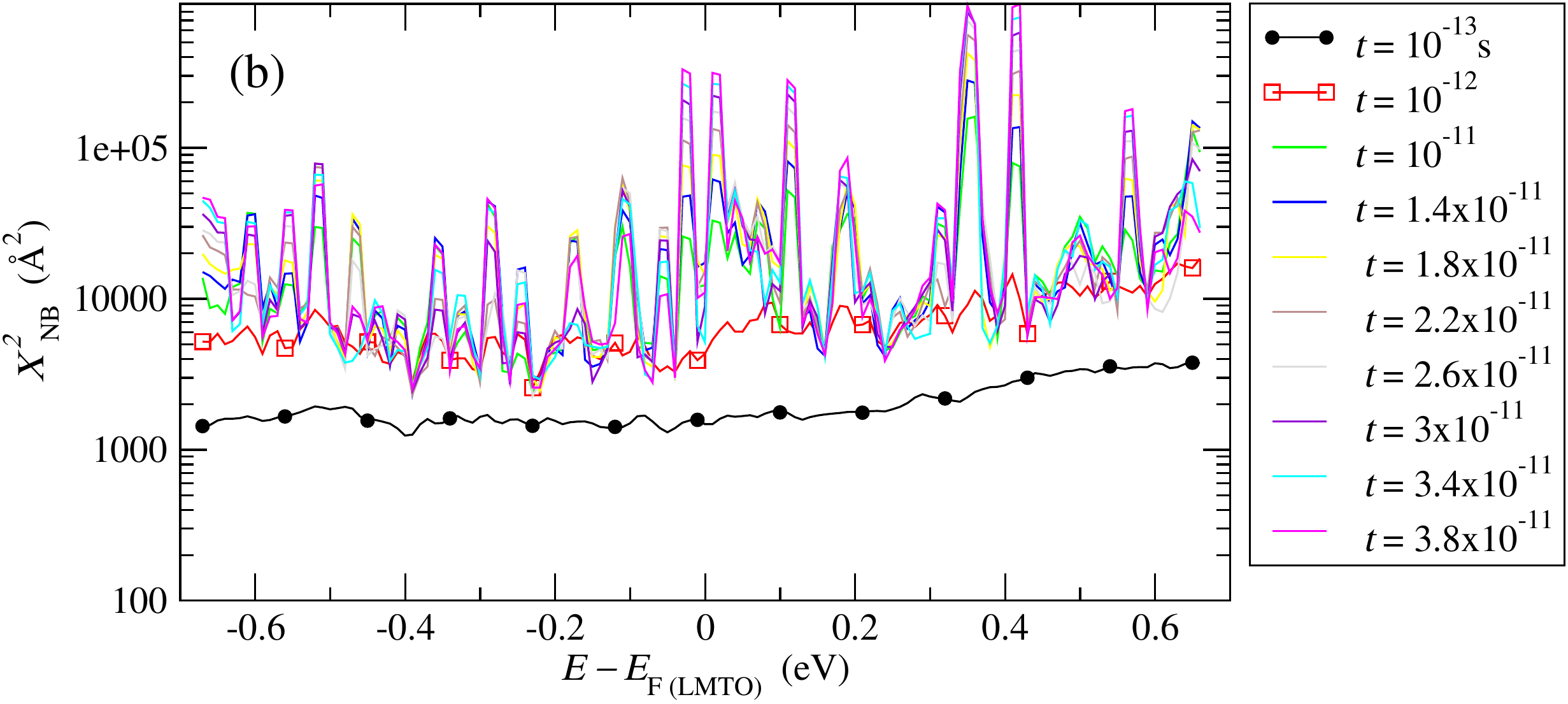}

\caption{\label{Fig_lambda_X2nB}
(color online) LMTO non-Boltzmann average square speading $X^2_{\rm NB}$   
in $\lambda$-Al$_{4.6}$Mn.
(a) $X^2_{\rm NB}$ versus time $t$ for various energy values,
(b) $X^2_{\rm NB}$ versus energy $E$ for various time values.
}
\end{center}
\end{figure*}

The mean square spreading is the sum of a quadratic term (Boltzmann term) and a non-Boltzmann term (\ref{Eq_DeltaX2}). 
Figure \ref{Fig_lambda_X2nB} presents the a typical behavior of the non-Boltzmann term $X_{\mathrm{NB}}$, versus
energy $E$ and time $t$. 
As expected from equation (\ref{Calcul_DeltaX2}), for large $t$ values $X_{\mathrm{NB}}(t)$ oscillates. 
For some energies the amplitude of oscillations is large, but for other energies this amplitude is small.
These last energies correspond approximatively to local minimum of the DOS.
Therefore they correspond to realistic values of the Fermi energy $E_{\rm F}$.
For these energies, at large $t$,  $X_{\mathrm{NB}}(t)$  is almost constant and one can 
define $L_{wp}(E)$ by,
\beq
 L_{wp}(E) \simeq \sqrt{X_{\rm NB}^2(E,t)}~~~{\rm for~large}~t.
\eeq
$L_{wp}(E)$ is the spatial extension of the wave packet at energy $E$.
From ab-initio calculations, its values varies from $\sim$50\,$\rm \AA$ to large values in $\lambda$-AlMn (figure \ref{Fig_lambda_X2nB}).
In $\alpha$-AlMnSi \cite{PRL06} and $1/1$-AlCuFe \cite{Trambly08} the minimum value of $L_{wp}(E)$ is about 20\,$\rm \AA$, which corresponds to the size
of smallest atomic clusters in these phases (Mackay clusters or Bergman clusters) \cite{Gratias00}. 
At these energies, one can then assume that,
\begin{equation}
 X^2(E,t) \simeq V_{\mathrm{B}}(E)^2 t^2 + L_{wp}^2(E).
\label{Eq_DeltaX2_QC}
\end{equation}
From ab initio calculations, this behavior is obtained for energy $E_{\rm F}$ corresponding the local minimum in the DOS as expected from stabilization mechanism.

\subsection{Conductivity in the Small Velocity Regime}

\begin{figure*}%
\begin{center}

%\includegraphics[height=5cm]{lambdaC_cond1-DOS.eps}
%\includegraphics[width=0.5\textwidth]{lambdaC_cond1-DOS.eps}
%
%\vskip .1cm
%\includegraphics[width=0.515\textwidth]{lambdaC_cond1-VB.eps}~~
%
%\vskip .1cm
\includegraphics[width=0.53\textwidth]{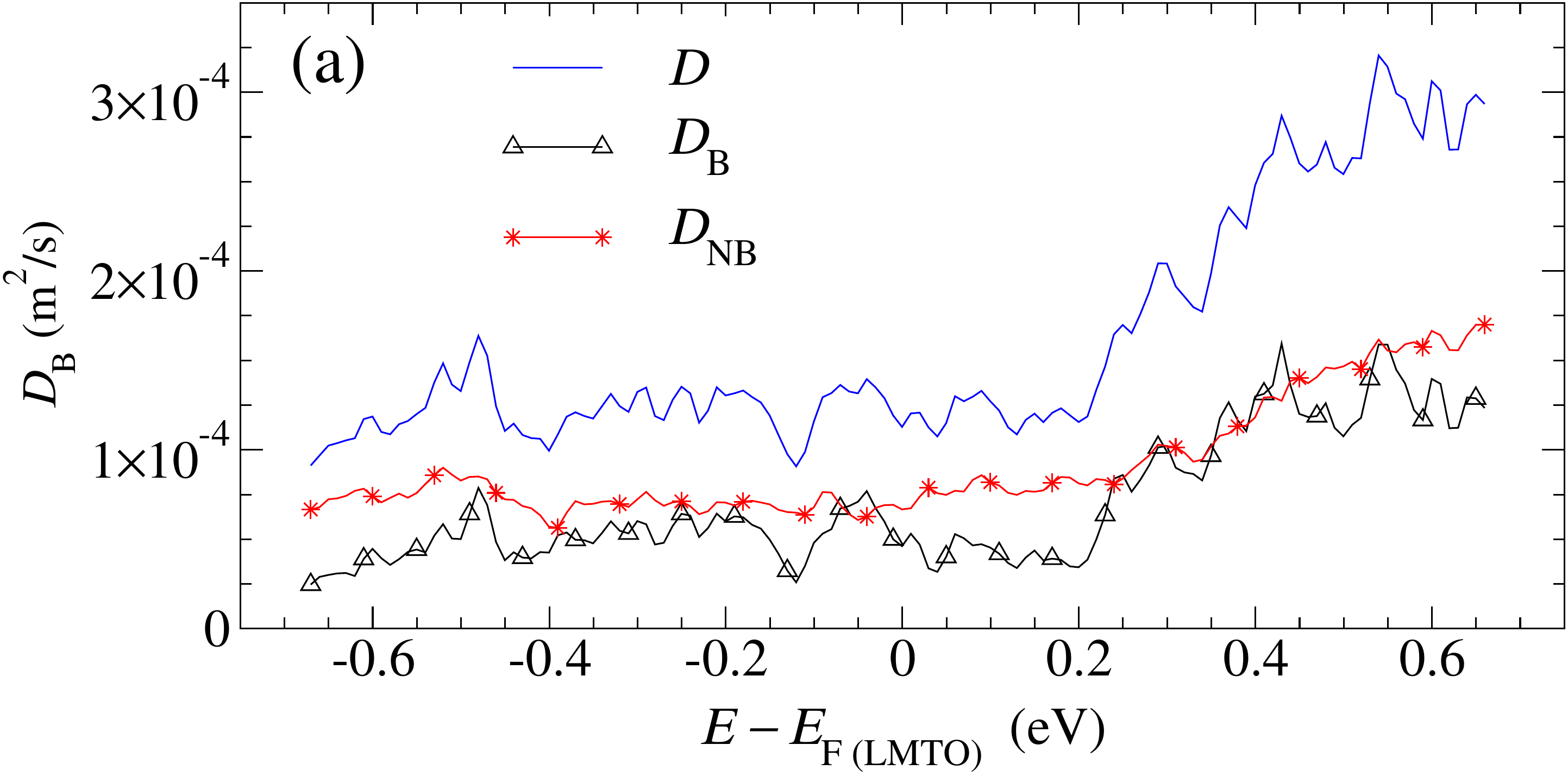}

\vskip .1cm
\includegraphics[width=0.53\textwidth]{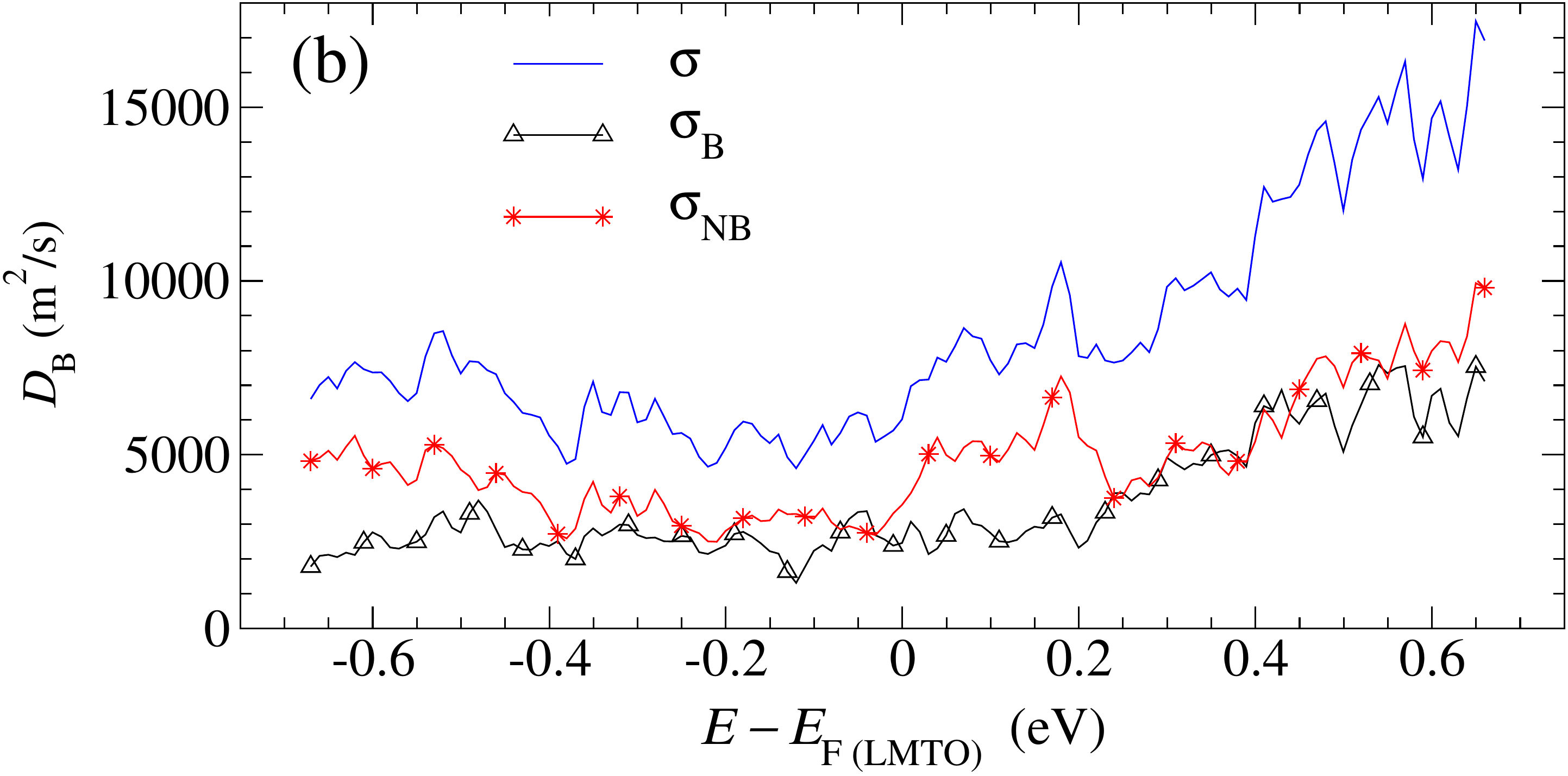}

\caption{\label{Figlambda_LMTO_D_sigma}
Electronic (a) diffusivity and (b) conductivity 
in $\lambda$-Al$_{4.6}$Mn calculated from ab-initio LMTO method 
for scattering time $\tau = 10^{-13}$\,s.
%Lines are guides for the eyes.
}
\end{center}
\end{figure*}

\begin{figure*}%[]
\begin{center}

\includegraphics[width=0.5\textwidth]{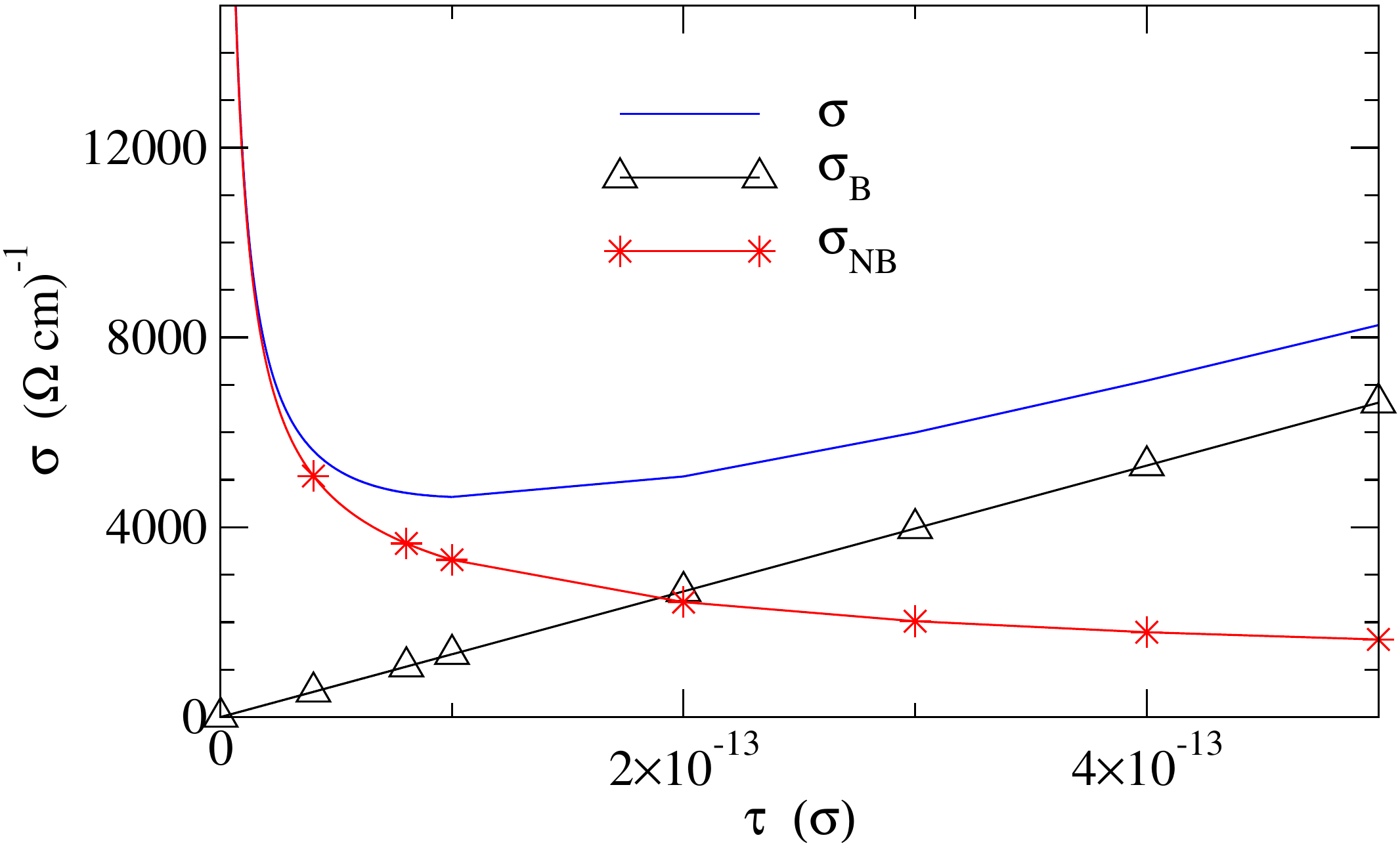}

\caption{\label{Fig_lambda_X2nB_fE}
LMTO Conductivity versus scattering time $\tau$ 
in $\lambda$-Al$_{4.6}$Mn at energy $E=-0.12$\,eV
}
\end{center}
\end{figure*}

\begin{figure*}%[]
\begin{center}

\includegraphics[width=0.45\textwidth]{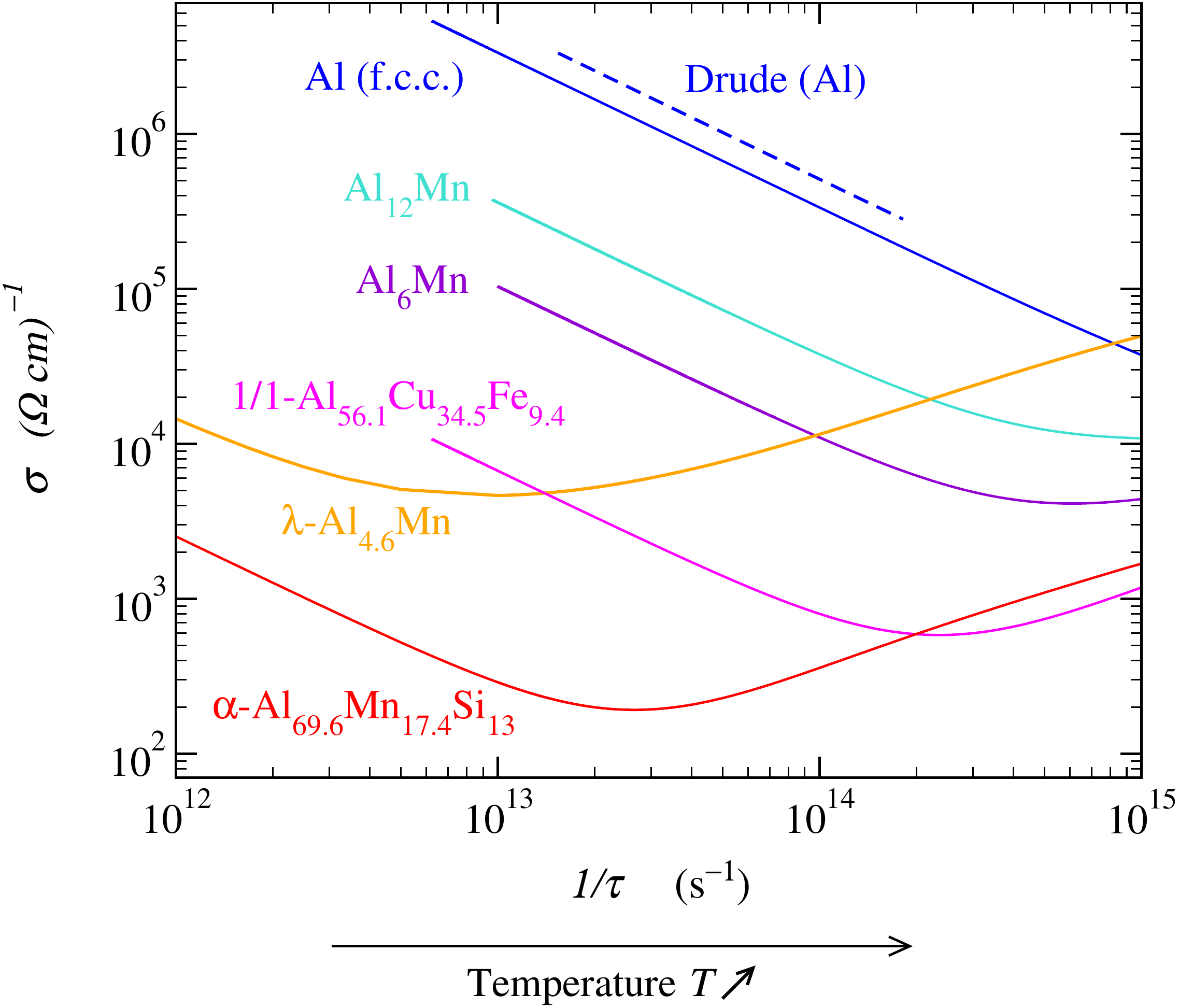}

\caption{\label{Fig_cond_loglog}
LMTO Conductivity $\sigma(E_m)$ versus inverse scattering time $1/\tau$ in Log-Log scale. 
For each phases $E_m$ is the energy near to $E_{\rm F}$ for which $\sigma(E)$ 
reaches a minimum value.
}
\end{center}
\end{figure*}

The semiclassical theory of transport in crystal is based on the concept of a charge carrier wave packet 
propagating at a velocity $V_{\rm B}$. 
Moreover in real materials, defects induce scattering events of the wave packet (elastic or inelastic scattering) separated by a average time $\tau$.
The validity of wave packet concept requires that the extension $L_{wp}$ of the wave packet is smaller that the distance $V_{\rm B} \tau$ of traveling betwen two scattering events. 
In phases with a small Boltzmann velocity and enough large extension of the wave packet this condition in no more valid and
\begin{equation}
 L_{wp} > V_{\rm B} \tau .
\label{Eq_SVR}
\end{equation}
Thus, when (\ref{Eq_SVR}) is satisfied
the semiclassical (Boltzmann) approach for transport is no more valid, and 
a new diffusion regime, called {\it ``Small Velocity Regime''} (SVR), is reached \cite{PRL06}.

For realistic values of scattering time, $\tau \simeq 10^{-14}$\,s or  $\tau \simeq 10^{-13}$\,s, in quasicrystals and approximants \cite{Mayou93}, our ab initio calculation show that the SVR is reached for many energies.
Results for $\alpha$-AlMnSi and $1/1$-AlCuFe are presented in Refs.\,\cite{PRL06,Trambly08}. 
The diffusivity and conductivity in  $\lambda$-AlMn are shown 
figures \ref{Figlambda_LMTO_D_sigma} and \ref{Fig_lambda_X2nB_fE}.

When (\ref{Eq_DeltaX2_QC}) is satisfied for realistic Fermi energy values, one obtains simple equations for the diffusivity 
\cite{PRL06},
\begin{equation}
D(E_{\rm F}) = V_{\rm B}^2(E_{\rm F}) \tau + \frac{1}{2} \frac{L^2_{wp}(E_{\rm F})}{\tau},
\label{Eq_SVR_D}
\end{equation}
and the conductivity,
\begin{equation}
\sigma(E_{\rm F}) = e^2 n(E_{\rm F})V_{\rm B}^2(E_{\rm F}) \tau + \frac{1}{2} e^2 n(E_{\rm F}) 
\frac{L^2_{wp}(E_{\rm F})}{\tau},
\label{Eq_SVR_sigma}
\end{equation}
where the first terms are the Boltzmann terms and the second terms the non-Boltzmann terms.
This two terms are shown in figures (\ref{Figlambda_LMTO_D_sigma}) and (\ref{Fig_lambda_X2nB_fE}) for $\lambda$ phase.
Figure \ref{Fig_cond_loglog} compares the conductivity in approximants and complex phases  
with simple phases that have a standard metallic behavior. 
From equation (\ref{Eq_SVR_D}), it is clear that Boltzmann (non-Boltzmann) term increases (decreases) when
$\tau$ increases. The minimum of diffusivity (conductivity) is thus obtained when
\be
\tau = \tau^* ~~{\rm with}~ \tau^* = \frac{L(E_{\rm F})}{\sqrt{2} \, V_{\rm B}(E_{\rm F})} .
\ee 
For a scattering time, $\tau > \tau^*$, the Boltzmann term dominates and the diffusivity (conductivity) increases as $\tau$ increases. As $\tau$ decreases when defects and/or temperature increase, the behavior
is thus {\it metallic like}: $\sigma$ decreases when defects and/or temperature increase.
But, for $\tau < \tau^*$, the conductivity increases when defects and/or temperature increase and 
the behavior is {\it insulating like}.
From ab initio calculations in realistic phases, $\tau^*$ is around a few $10^{-14}$ or $\sim 10^{-13}$\,s. 
Theses scattering time values correspond to scattering time estimates in quasicrystals and approximants from transport measurements at low temperature (4\,K) \cite{Poon92,Berger93,Mayou93}. Therefore, when temperature increases from low temperature the behavior of these complex phases is insulating like as found experimentally.
From equation (\ref{Eq_SVR_sigma}) when Boltzmann term is negligeable, $\tau \ll \tau^*$, the conductivity follows the {\it inverse Mathiessen rule} found experimentally \cite{Berger93,Mayou93}:
\be
\sigma(T) = \sigma_{\rm 4 K} + \Delta \sigma(T)
\ee

In $\alpha$-AlMnSi the minimum value of the conductivty obtained from ab initio calculation, $\sigma(E_{\rm F},\tau^*)$, is about 200\,$\rm (\Omega cm)^{-1}$ which is in good agreement with measurements \cite{Berger93}.
This value is very low with respect to standard metallic alloys (figure \ref{Fig_cond_loglog}) as 
expected in Al-based quasicrystals.
In the complex metallic alloys $\lambda$-AlMn, the minimum value of  $\sigma(E_{\rm F},\tau^*)$ is not so low, but the insulating like regime is obtained for a larger range of $\tau$ values as illustrated on figure (\ref{Fig_cond_loglog}). This shows that the small velocity regime can be observed in a great number of 
complex metallic alloys even if their conductivity is not very low. Indeed, Dolin$\rm \check{s}$ek et al. \cite{Dolinsek07,Dolinsek09} (see also the review \cite{Dolinsek12} and Refs. therein)
were able to analyze experimental transport properties of several complex metallic phases by using the small velocity regime model.  

It must be noted that the quick variation of the DOS with energy implies also that the DOS will be modified by disorder. This also can contribute to the variation of the conductivity. Indeed the discussion here focuses on the variation of the diffusivity but a variation of the DOS also contributes to a variation of the conductivity.  Yet we believe that the variation of the diffusivity is an important ingredient as indicated by the numerical values obtained in this model. In addition as explained in Ref. \cite{PRL06} the small velocity regime also explains the absence of a Drude peak in the low frequency optical conductivity that is observed experimentally.

\section{Metal-Insulator transition}
\label{SecMIT}

Let us discuss now the nature of the phase at zero temperature as a function of the static disorder \cite{Lee85}. We recall here that we consider only non interacting electrons in a three dimensional system. For standard metals it is well known that static disorder can induce a transition from a metallic to an insulating state when disorder increases. This is the Anderson transition.  Here we discuss the role of static disorder for the case where the electrons propagate in an unusual way with a Non-Boltzmann contribution to diffusion that cannot be ignored. As we show this may strongly modify the occurrence of the insulating state. We discuss  the metal-insulator phase diagram at zero temperature  according to the scaling theory of localization.  According to this  theory, a central quantity is the conductance $g$ of a cube with a 
size equal to the elastic mean free path $ X(E_{\mathrm{F}},\tau)$,
\ben
g \simeq e^2 n(E_{\mathrm{F}}) D(E_{\mathrm{F}},\tau)  X(E_{\mathrm{F}},\tau) ,
\een
where $n(E_{\mathrm{F}}) $  is the density of states at the Fermi energy and $D(E_{\mathrm{F}},\tau) $ is the diffusivity computed in the Relaxation Time Approximation approximation.  The typical propagation length $ X(E_{\mathrm{F}},\tau)$ on a time scale $\tau$, i.e. the mean-free path, is such that:
\ben 
 X^2(E_{\mathrm{F}},\tau)=  X_{\NB}^2 +  V^2 \tau^2 .
\een

Let us introduce $g_0$, which is characteristic of the {\it perfect crystal} and is defined by:

\ben 
g_0 = e^2 n(E_{\mathrm{F}})   X_{\NB}^2 V
\label{g0}
\een

Let us introduce an adimensional  value $\tilde{\tau}$ of the scattering time $\tau$ defined by,
\ben 
\tilde{\tau} = \frac{V \tau}{ X_{\NB} } = \frac{\tau}{\sqrt{2} \tau^*} .
\label{tildetau}
\een 
Let us define also  the function $f(x)$,
\ben
f(x)= \left(\frac{1}{2 x} + x \right) \sqrt{(1 + x^2)} ,
\label{f(x)}
\een
then one has,
\ben
g= g_0 f(\tilde{\tau}) .
\een

\begin{figure*}%[]
\begin{center}

\includegraphics[width=0.45\textwidth]{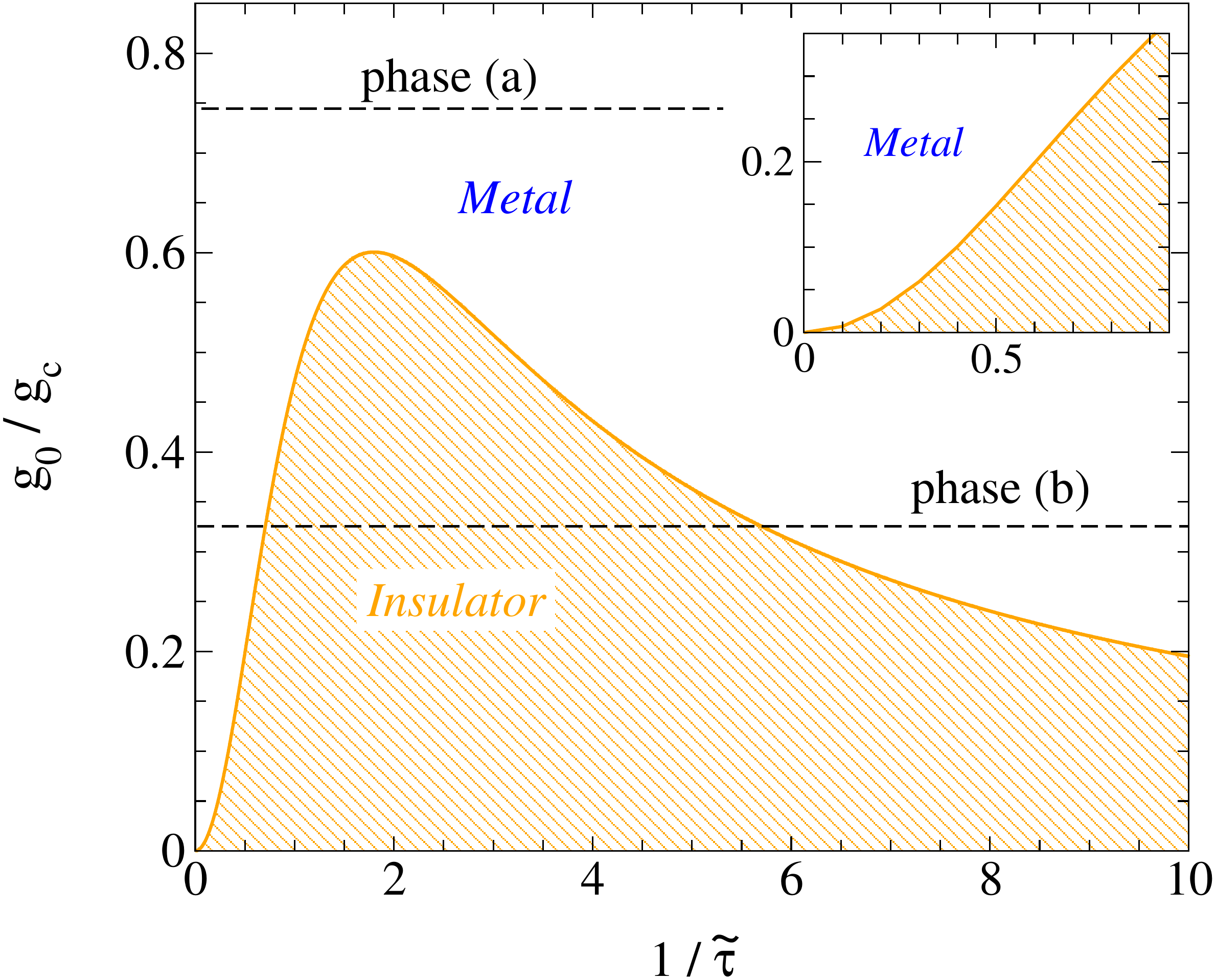}

\caption{\label{Fig_MI_transition}
Metal-Insulator phase diagram as a function of the two parameters 
$g_0/g_c$ and $1/\tilde{\tau}={\sqrt{2}\tau}/{\tau^*}$. 
The insert represents the limit of a normal metal i.e. for  fixed 
$\tau$ and $V$ the limit of a small $ X_{\NB}$. 
After ($\ref{g0}$) and ($\ref{tildetau}$) this limit is 
in the region of the phase diagram at small $g_0/g_c$ and small $1/\tilde{\tau}$.
}
\end{center}
\end{figure*}

After the scaling theory \cite{Lee85} a three dimensional  system is insulating 
(metallic, respectively)  if $g<g_c$ (resp. $g>g_c$) where $g_c$ is the value of the universal critical 
conductance in the scaling theory. Using $g= g_0 f(\tilde{\tau})$ it is 
equivalent to say that the system is insulator if  $g_0/g_c<1/f(\tilde{\tau})$ and metallic if $g_0/g_c>1/f(\tilde{\tau})$. We emphasize that $g_0/g_c$  is characteristic of the perfect crystal 
whereas $1/\tilde{\tau}$ measures the scattering   rate $1/\tau$ in units of  $V/ X_{\NB}$. The figure  \ref{Fig_MI_transition} illustrates this phase diagram.

A first remarkable property of this phase diagram is that if $g_0> R g_c$  with $R=2/(\Phi)^{5/2}$ where $\Phi$ is the Golden Mean ($R\simeq 0.6$)  then the system is always metallic whatever 
the value of the scattering rate  (phase (a) in figure \ref{Fig_MI_transition}).  This is not the case for normal  metals that always become insulating at sufficiently small scattering time 
$\tau$ (i.e. at  sufficiently large disorder). Note that for a system like AlMnSi $g_0/ g_c \simeq 2$ and therefore this phase should always be metallic independently of the amount of disorder.

If  $g_0<Rg_c$ the system is metallic at large and small scattering rates  and insulator in an intermediate zone
(phase (b) in figure \ref{Fig_MI_transition}). This means that if the system is in the large $1/\tilde{\tau}$ metallic 
region it will become insulating by {\it decreasing} $1/\tilde{\tau}$ that is by  {\it decreasing} disorder! This is just the opposite of the standard conditions 
for the occurrence of the Anderson localization transition. This anomalous behavior occurs because in that regime the quantum diffusion is dominated by the non-Boltzmann term and not by the ballistic term.  The other insulator-metal transition is normal in the sense that the metallic state is obtained by decreasing disorder. 

Note that the case of a normal metal corresponds to 
the limit $ X_{\NB} \to 0$. 
In that case one uses the asymptotic form of the function $f(\tilde{\tau})$ 
for large $\tilde{\tau}$  namely $f(\tilde{\tau})\simeq \tilde{\tau}^{2}$.
One then recovers the standard criterion for free-like electrons.

\section{Conclusion}
\label{Conclusion}

To summarize this article shows that approximant phases $\alpha$-AlMnSi, $1/1$-AlCuFe and the complex phase $\lambda$-AlMn present unusual band structure and Bloch states. This can explain their anomalous transport properties when compared to standard metallic phases. In particular the analysis of the quantum diffusion in these phases shows that it is badly reproduced by the standard semi-classical theory. As we find the square of the quantum diffusion length  is the sum of two terms that depend on time. One term is the ballistic contribution and the other term is the Non-Boltzmann contribution. Depending on the scattering time one term or the other can dominate the conductivity. If the ballistic term dominates this corresponds to a standard metallic behaviour. If the non-Boltzmann term dominates (small velocity regime)  this induces an insulating like behaviour that is in good agreement qualitatively and even quantitatively with the experimental results. As we discussed also the occurrence of an Anderson transition is also deeply affected by the anomalous quantum diffusion and the possible existence of a small velocity regime. 

We note also that a small velocity regime can be found in other systems that present flat electronic bands. This is the case in the recently studied rotated bilayers of graphene \cite{Graphene10,Graphene12,Brihuega12}. Indeed these systems have large unit cells and it has been shown that the electronic coupling between the two layers tends to decrease the Fermi velocity and even cancel it at some specific small angles.

\section*{Acknowledgments}

The computations have been performed at the
Centre de Calcul (C.D.C),
Universit\'e de Cergy-Pontoise.
Part of the numerical results has been
obtained by using  the Condor Project
(http:$/\!/$www.condorproject.org$/$).
I thank also Y. Costes and D. Doumerge, C.D.C.,
for computing assistance.

\end{document}